\documentclass[conference]{IEEEtran}

\usepackage{graphicx}
\usepackage{booktabs}
\usepackage{cite}
\usepackage{url}
\usepackage{amsmath,amssymb,amsfonts}
\usepackage{xcolor}

\usepackage{caption}
\graphicspath{{./}}

\begin{document}
\title{
D-TINs for Track-Sampled Ice Sheet Surface Reconstruction from Satellite Altimetry Data}

\author{
\IEEEauthorblockN{Xiaoyu Xia}
\IEEEauthorblockA{Department of Computer Science\\
Tufts University\\
Medford, MA, USA\\
xiaoyu.xia@tufts.edu}
\and
\IEEEauthorblockN{Noah Bergam}
\IEEEauthorblockA{Department of Computer Science\\
Columbia University\\
New York, NY, USA\\
njb2154@columbia.edu}
\and
\IEEEauthorblockN{Abani Patra}
\IEEEauthorblockA{Department of Computer Science\\
Tufts University\\
Medford, MA, USA\\
abani.patra@tufts.edu}
}

\maketitle

\begin{abstract}
Satellite altimetry samples ice sheets densely along tracks but sparsely across tracks, with frequent gaps in spatial coverage. We study  the impact of this irregular support on surface reconstruction and introduce a new method, D-TIN, that applies Delaunay triangulation and barycentric interpolation directly to observed locations. A spatial holdout removes contiguous track intervals and nearby same-pass observations. We compare D-TIN with two common workflows, kriging and tile-based polynomial interpolation, to evaluate its accuracy and computational efficiency. Across seven ICESat-1 campaigns, D-TIN reduces RMSE by 12.1~m to 23.6~m relative to independent 5~km tile and polynomial based reconstruction. Its aggregate accuracy is close to the ordinary kriging workflow using a 1~km grid on the commonly used GStatSim platform  while its median  runtime is 252 times lower. 
A fixed-weight multiscale extension of D-TIN, called MS-D-TIN, gives the lowest aggregate RMSE, 67.5~m, while retaining the low computational cost of the D-TIN family.

\end{abstract}


\section{Introduction}
The Greenland Ice Sheet holds an ice volume equivalent to about 7.4~m of global mean sea-level rise\cite{morlighem2017} and is losing mass at an accelerating rate with potentially severe consequences \cite{bamber2018,imbie2020}. Satellite laser altimetry measures surface elevation along satellite ground tracks and is central to monitoring ice sheet change at a continental scale\cite{zwally2002}. 

The resulting track-based sampling geometry is strongly anisotropic and contains gaps. For example, the ICESat-1 measurements are about 172~m apart along a ground track, while adjacent tracks can be separated by kilometers \cite{schutz2005}.
The observations therefore form an irregular, track-sampled spatial support rather than a regular grid. Quality filtering also leaves additional spatial gaps within some observed sequences. Surface reconstruction must estimate elevations across gaps between and within these tracks.

Two common workflows impose a spatial structure before interpolation. Ice sheet DEM pipelines aggregate measurements into regular cells and use spatial fitting or kriging to produce a raster \cite{helm2014,slater2018}. GStatSim provides a gridded-kriging workflow for irregular observations \cite{mackie2023}. Patch-based pipelines fit a low-order polynomial surface inside predetermined tiles \cite{serac2012,csatho2014}. These structures are convenient for computation and data fusion, but they change the support available to the estimator. Grid based representations move interpolation anchors to cell locations, while fixed tiles separate measurements that lie across a tile boundary.

 This paper proposes and evaluates a Delaunay triangulated irregular network, or D-TIN, and a multiscale variant(MS-D-TIN). 
 D-TIN uses the measurement locations as vertices, making direct use of the measurement data. 
 In this approach, the mesh (triangulation) remains dense along the tracks and connects observations across track and tile boundaries efficiently. Prediction uses barycentric interpolation inside each triangle. 
 Delaunay triangulation maximizes the minimum angle among triangulations of a point set \cite{lawson1977} and minimizes the integrated squared gradient of the corresponding piecewise-linear interpolant \cite{rippa1990}. The construction is deterministic and has no fitted covariance or representation scale. 
These properties favor a geometrically regular piecewise-linear interpolation while retaining the observed measurement locations as support; long triangles can still incur large errors over rough terrain.

This work makes three contributions. First, D-TIN is evaluated against independent tile polynomials and a GStatSim gridded-kriging workflow. D-TIN is more accurate than the tile method in all seven campaigns
with the largest margins in steep terrain and across wide gaps. It has overall accuracy close to the evaluated kriging workflow at much lower computational cost. Second, a rigorous spatial holdout testing strategy is developed to remove contiguous track intervals and nearby same-pass observations from training, providing more reliable test error estimates. 
Third, controlled changes to support placement and connectivity identify the effects of tile boundaries and grid aggregation. We also develop a fixed-weight multiscale D-TIN as an extension of the D-TIN with potential improvement in accuracy.

\section{Related Work}
Kriging is the best linear unbiased predictor when its covariance structure is known\cite{cressie1993}. In practical applications, the covariance structure and its fitting scale must be estimated, introducing additional estimation error and high computational costs. Ice sheet elevation workflows usually aggregate altimetry to a regular grid and use spatial fitting or kriging for interpolation \cite{helm2014,slater2018}. GStatSim packages this gridded-kriging procedure into a public geostatistical workflow \cite{mackie2023}. 
We evaluate this complete workflow and use raw-support kriging, that is, the same kriging applied directly to the original measurement locations instead of the gridded cell averages,  to measure the effect of its gridding stage.

Patch-based methods provide another representation. For example, the Surface Elevation Reconstruction And Change method (SERAC) \cite{serac2012} fits polynomial surfaces around track crossover regions to estimate elevation change from repeated campaigns. Schenk and Csath\'o \cite{serac2012} show that near-parallel repeat tracks do not independently resolve the cross-track slope and elevation in a local polynomial fit and crossover geometry or external slope information can provide the missing constraint. SERAC and the Approximation
by Localized Penalized Splines framework (ALPS)  use these estimates for multi-epoch ice sheet analysis \cite{csatho2014, alps2021}. 
Earlier digital elevation model (DEM) studies compared TIN and kriging on field surveys and contour data, with results that varied by terrain and sampling density \cite{chaplot2006}. ICESat studies have also compared methods for estimating elevation change along repeat tracks \cite{felikson2017}. 

Unstructured triangulations are also used as finite-element meshes in ice-sheet models \cite{larour2012}. The triangulation based reconstructions advocated here can therefore provide simple to use boundary conditions for such analyses.

\section{Methods}
\label{sec:methods}
\subsection{D-TIN with Barycentric Interpolation}

\begin{figure}[t]
\centering
\includegraphics[width=0.8\columnwidth]{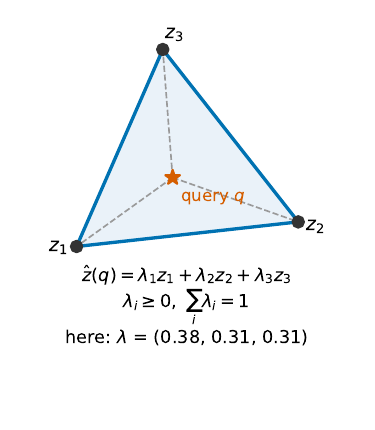}
\caption{
Barycentric interpolation inside one Delaunay triangle. The weights are area ratios determined by the geometry.}
\label{fig:method}
\end{figure}

Given training observations $\{(x_i, y_i, z_i)\}_{i=1}^{N}$, we construct the Delaunay triangulation of the horizontal coordinates. A query $q$ inside a triangle is predicted by
\begin{equation}
\hat z(q) = \lambda_1 z_1 + \lambda_2 z_2 + \lambda_3 z_3,\qquad
\lambda_i \ge 0,\ \textstyle\sum_i \lambda_i = 1,
\end{equation}
where the $\lambda_i$ are barycentric coordinates (Fig.~\ref{fig:method}).
For $N$ training observations, a planar Delaunay triangulation has $O(N)$ storage and can be constructed in $O(N\log N)$ time \cite{shewchuk1996}. We use the SciPy implementation for triangulation and query location \cite{virtanen2020}. No grid resolution, neighbourhood size, or covariance parameter is fitted. 

The barycentric weights depend only on the query and the three vertices of its containing simplex. So prediction is a convex combination of observed elevations. It cannot overshoot the local vertex range. Long triangles can still bridge rough terrain, so geometric support does not remove error at larger support distances. The triangulation determines which observations define the local plane. The evaluation keeps test observations within the training convex hull.

Construction and prediction are separate operations. Triangulation creates a sparse planar graph with fewer than $3N$ edges. Query location identifies one simplex, after which interpolation uses three values and constant work. This matters for repeated reconstruction because the mesh can be reused for many query sets as long as the training support is unchanged.

\subsection{Tile Polynomials}
The tile baseline partitions the domain into 5~km squares and fits an independent polynomial of degree up to three per tile \cite{serac2012}. Observations within a tile are often nearly collinear, so higher-order terms may be constrained along track but unidentified across track. 

A polynomial order is admitted only when the smallest and largest singular values of the scaled design matrix satisfy $\sigma_{\min}/\sigma_{\max} > 10^{-2}$, ensuring that the measurements adequately span the tile in the required direction. Otherwise, the model backs off to a lower degree. A fitted tile uses only observations from the query's tile. If no model can be fitted there, prediction uses the globally nearest training observation so that every test point remains in the evaluation.

Each tile can be processed independently, and polynomial coefficients are easy to store. It also imposes a hard access boundary. An observation just outside a tile cannot affect a query inside it. 

\subsection{Gridded Kriging}
The geostatistical baseline uses GStatSim to average training observations into 1~km cells and then apply simple and ordinary kriging \cite{mackie2023}. 
 It applies a normal-score transform and uses an isotropic exponential variogram. Variogram fitting uses a seeded subsample of at most 5{,}000 occupied cells. Prediction uses up to 100 neighbors within 100~km. The search radius increases to 200~km only when a query has too few neighbors. Predictions are then transformed back to elevation.

We vary the variogram fitting window, \texttt{maxlag} from 25 to 1600~km. The main comparison uses ordinary kriging at 800~km (OK800). Simple kriging at 50 and 800~km is retained for the sensitivity analysis.
Each prediction solves a dense system over at most $n$ neighbors, so the prediction stage costs $O(Pn^{3})$ for $P$ query points. With neighborhood size $n=100$ here, this is on the order of $10^{6}$ operations per query. We therefore rely on paired end-to-end runtimes for cost comparison.

\begin{figure*}[t]
\centering
\includegraphics[width=0.94\textwidth]{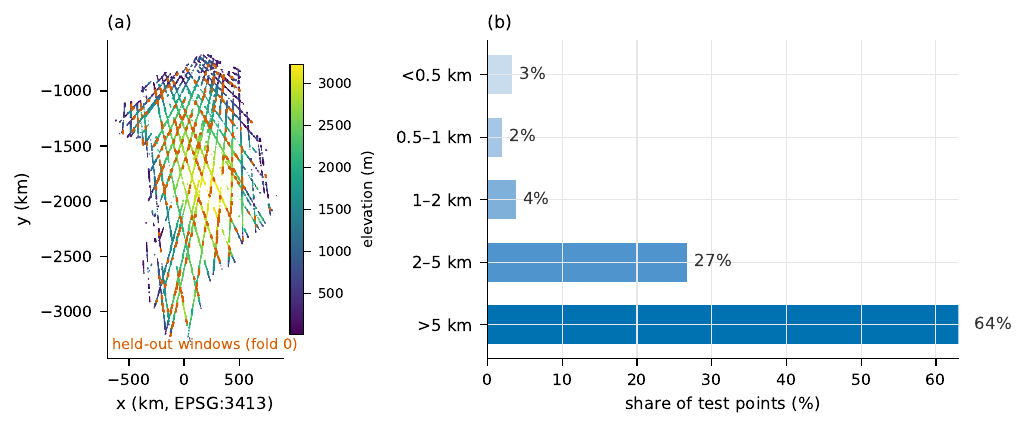}
\caption{Evaluation geometry. (a) Held-out data is shown in red. (b) Distance from each test point to the nearest training measurement. In 2005, 64\% of test points are more than 5~km away and about 5\% are within 1~km.}
\label{fig:setup}
\end{figure*}

 \subsection{Multiscale D-TIN }

D-TIN supports a simple multiscale extension because the same deterministic construction can be applied at several support resolutions. We build  D-TINs from cell-mean support at 1, 5, and 10~km and combine their predictions with fixed weights 0.6, 0.3, and 0.1. This variant is named  MS-D-TIN. The construction follows the general use of successively coarser approximation spaces for multiscale data reduction \cite{shekharpatra2020}. The current weights are empirically chosen and fixed for this study. Systematic multiscale weight selection is ongoing work. MS-D-TIN tests whether combining several support scales can improve aggregate error within the same method branch.

\section{Data and Evaluation Protocol}
\label{sec:protocol}

\subsection{ICESat-1 Observations}

We use Greenland ice sheet elevations derived from the ICESat-1 GLAS GLA12 Version~34 product \cite{zwally2014gla12} available at the National Snow and Ice Data Center(NSIDC). 
Elevations are referenced to the WGS84 ellipsoid. Horizontal coordinates are projected to EPSG:3413, and returns below 5~m elevation are removed. The main evaluation contains six monthly batches from three 2005 GLAS laser campaigns \cite{schutz2005}, L3B, L3C, and L3D (February--March, May--June, and October--November).
Four campaigns from 2003, 2004, and 2006 provide additional replications (L1A, L2A, L3A, and L3E). 
The six monthly batches from 2005 are divided into folds for the detailed error and runtime analysis (Table~\ref{tab:main}). The seven campaigns are distinct acquisition periods spanning 2003--2006 and provide the units for the cross-campaign comparison (Table~\ref{tab:campaigns}).

\subsection{Spatially Structured Train--Test Splits}

A random point-wise split  usually leaves measurements
immediately adjacent to a held-out point on the same ground
track in the training set. The reported error would then primarily reflect interpolation between neighboring footprints rather than reconstruction across a missing track interval. To avoid this, we develop a spatial splitting strategy that holds out contiguous along-track windows and removes nearby observations from the same satellite track pass, following spatial cross-validation principles \cite{roberts2017,ploton2020}.  

Each test window contains a contiguous 10-35~km interval of a satellite track pass, with at least 30 observed elevations. We also exclude training observations within 2 km of either end of the window along the same pass. The buffer prevents the methods from reconstructing the window using immediately adjacent along-track measurements while leaving observations on neighboring tracks available.

Quality filtering can create gaps within a continuous satellite pass. If the 2 km buffer ended at the gap, an observation from the same pass lying just beyond it could remain in the training set. We therefore define continuous pass segments from consecutive footprints along the track, treating a break as the start of a new segment only when the along-track separation exceeds 50 km. This threshold is much larger than the gaps introduced by quality filtering, so it distinguishes separate overpasses without splitting a single one. The test window and its 2 km buffer are removed from the same segment. Test points are additionally required to lie within the convex hull of the remaining training observations, so every evaluated prediction is an interpolation rather than an extrapolation. Since the method uses local interpolation, this restriction to the convex hull is justified.

Eligible windows are assigned to five folds within each of the six 2005 monthly batches, giving 30 paired folds in total. Every method receives the same training and test observations. The prediction target is the recorded elevation at each held-out GLAS footprint. The reported error therefore measures how well missing observations are reconstructed, not how closely the output matches an independently observed continuous surface. For the cross-campaign comparison, residuals are pooled within each campaign, so every held-out observation from that acquisition period contributes equally.

\subsection{Metrics and Stratification}
 We report root mean squared error (RMSE), mean absolute error (MAE), and median absolute error (MedAE). MedAE provides a robust summary when elevation residuals are non-Gaussian \cite{hohle2009}. 
All errors are reported in meters. In data from 2005, 64\% of test points are more than 5~km from the nearest training measurement, and only about 5\% are within 1~km (Fig.~\ref{fig:setup}).

Terrain strata are computed from training observations only. For each test point, we compute the gradient magnitude of the planar surface on its containing training triangle. Low, moderate, and high slope correspond to values below 0.005, from 0.005 to 0.02, and above 0.02. Distance strata use the Euclidean distance to the nearest training observation. 

RMSE emphasizes large residuals and is the primary aggregate measure used by the compared surface workflows.
MAE gives a linear penalty, while MedAE describes a typical prediction under a heavy-tailed residual distribution. None of these metrics alone identifies where an error occurs.

Table~\ref{tab:main} reports the mean of the 30 fold-level accuracy metrics and the median fold runtime. Table~\ref{tab:terrain} and Table~\ref{tab:mechanism} aggregate the cross-validated residuals by slope and distance before computing RMSE. Table~\ref{tab:campaigns} shows the comparison across seven campaigns, 
pooling residuals within each campaign before calculating each metric. 

\subsection{Experimental Configuration}
All accuracy comparisons use identical fold configurations. Stochastic components, including the variogram subsample, use fixed seeds. 
All methods run single-threaded.
The runs on 2005 data execute on Intel Xeon Gold 5220R and 6226R nodes with 48GB of memory allocated per single-core
task. A paired fold places the compared methods on the same node, and timing covers representation construction, model fitting, and prediction.

\section{Results}
\label{sec:results}

\begin{table}[t]
\caption{
Reconstruction error and runtime over 30 folds from 2005. Errors are means of fold-level metrics in meters. Runtime is the median across folds. SK and OK denote simple and ordinary kriging, and the suﬀix gives the variogram maxlag in kilometers.}

\label{tab:main}
\centering
\footnotesize
\begin{tabular}{@{}lrrrr@{}}
\toprule
Method & RMSE / m & MAE / m & MedAE / m & Time / s \\
\midrule
Tile polynomial, 5 km & 89.5 & 39.0 & 15.0 & 2.3 \\
SK50 & 117.3 & 63.6 & 32.0 & 385.3 \\
SK800 & 67.9 & 22.8 & 5.1 & 390.4 \\
OK800 & 67.7 & \textbf{22.6} & 5.1 & 398.2 \\
D-TIN & 68.8 & 22.8 & \textbf{4.3} & 1.5 \\
MS-D-TIN & \textbf{67.5} & 23.0 & 4.6 & \textbf{1.1} \\
\bottomrule
\end{tabular}
\end{table}

\begin{table}[t]
\caption{
RMSE by slope and distance strata over 30 folds in 2005. Distance is measured to the nearest training observation. Differences are competitor minus D-TIN.}
\label{tab:terrain}
\centering
\footnotesize
\begin{tabular}{@{}llrrrr@{}}
\toprule
Terrain & Dist. & \multicolumn{3}{c}{RMSE / m} & $n$ \\
\cmidrule(lr){3-5}
& & D-TIN & OK800$-$D & Tile$-$D & \\
\midrule
low slope     & $<$1 km & 2.3 & 0 & $+2.7$ & 33{,}873 \\
              & 1--5 km & 12.7 & $-0.7$ & $+1.8$ & 220{,}646 \\
              & $>$5 km & 23.5 & $-1.0$ & $+9.7$ & 489{,}485 \\
moderate slope& $<$1 km & 6.8 & 0 & $+8.9$ & 22{,}919 \\
              & 1--5 km & 43.7 & $-2.7$ & $+4.5$ & 121{,}671 \\
              & $>$5 km & 82.6 &$-2.8$ & $+32.1$ & 257{,}453 \\
high slope    & $<$1 km & 27.3 & $+3.1$ & $+35.1$ & 8{,}695 \\
              & 1--5 km & 144.1 & $-1.9$ & $+18.7$ & 36{,}877 \\
              & $>$5 km & 237.5 & $-1.7$ & $+62.9$ & 51{,}008 \\
\bottomrule
\end{tabular}
\end{table}

\begin{table}[t]
\caption{Reconstruction errors by campaign. Residuals are pooled within each campaign.  OK800 is ordinary kriging with an 800~km maxlag. Tile RMSE is higher than  D-TIN RMSE in all seven campaigns, while  D-TIN MedAE is lower than OK800 MedAE in all seven.}

\label{tab:campaigns}
\centering
\footnotesize
\begin{tabular}{@{}lrrrrrr@{}}
\toprule
 &  & \multicolumn{3}{c}{RMSE / m} & \multicolumn{2}{c}{MedAE / m} \\
\cmidrule(lr){3-5}\cmidrule(lr){6-7}
Campaign & Test $n$ & D-TIN & Tile & OK800 & D-TIN & OK800 \\
\midrule
L1A & 429,205 & 22.8 & 34.9 & 23.2 & 0.7 & 1.0 \\
L2A & 504,329 & 27.3 & 44.0 & 27.2 & 2.9 & 3.3 \\
L3A & 454,446 & 79.3 & 102.9 & 78.7 & 4.4 & 4.9 \\
L3B & 447,543 & 82.4 & 102.1 & 79.9 & 4.7 & 5.4 \\
L3C & 415,599 & 54.0 & 72.0 & 53.2 & 3.6 & 4.1 \\
L3D & 379,485 & 66.8 & 89.7 & 66.7 & 3.8 & 4.6 \\
L3E & 192,654 & 58.3 & 75.4 & 57.0 & 3.0 & 3.4 \\
\bottomrule
\end{tabular}
\end{table}

\subsection{Reconstruction Accuracy}

D-TIN has lower error than the fixed tile construction in every campaign (Table~\ref{tab:campaigns}). The RMSE is 20.7~m lower, and its MedAE is 10.7~m lower in the 2005 data(Table~\ref{tab:main}). 
Campaign difficulty varies widely. D-TIN RMSE ranges from 22.8~m for L1A to 82.4~m for L3B. D-TIN reduces
RMSE relative to the tile baseline by 12.1 to 23.6~m across the seven campaigns(Table~\ref{tab:campaigns}).

The slope and distance breakdown shows where the difference lies. The tile method has higher RMSE in every setting in Table~\ref{tab:terrain}. The gap is 2.7~m in low-slope terrain within 1~km of training support, and it increases to 35.1~m in nearby high-slope terrain and 62.9~m beyond 5~km. In general, fixed tiles lose accuracy both in local reconstruction and
prediction across wide gaps.

\begin{figure}[t]
\centering
\includegraphics[width=\columnwidth]{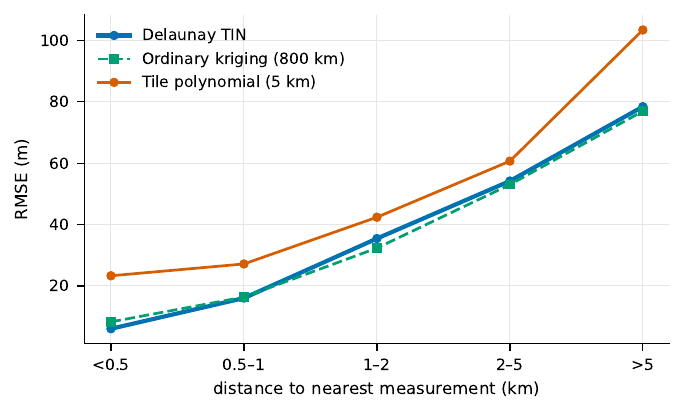}
\caption{
RMSE by distance to the nearest training observation. D-TIN has lower RMSE than the tile polynomial, while ordinary kriging at 800~km has a small advantage at larger distances from training support.}
\label{fig:strata}
\end{figure}

The aggregate accuracy of OK800 and D-TIN is close. In Table~\ref{tab:main}, the RMSE of OK800 is 1.1~m lower, and its MAE is 0.2~m lower, while D-TIN lowers MedAE by 0.8~m. Also, D-TIN has lower MedAE in every campaign with a median difference of 0.5~m (Table~\ref{tab:campaigns}). 
Fig.~\ref{fig:strata} shows that OK800 has a slight advantage beyond 1~km. D-TIN matches it in the low- and moderate-slope near field and lowers RMSE by 3.1~m in nearby high-slope terrain (Table~\ref{tab:terrain}).

MS-D-TIN has the lowest aggregate RMSE in Table~\ref{tab:main}. Its RMSE is 67.5~m, 1.3~m below D-TIN and 0.2~m below OK800. Its MedAE is 4.6~m, compared with 4.3~m for D-TIN. 
The added scales appear to modestly improve the tail-sensitive aggregate metric while giving up some typical local accuracy.

\subsection{Parameter Sensitivity}
\begin{figure*}[t]
\centering
\includegraphics[width=0.86\textwidth]{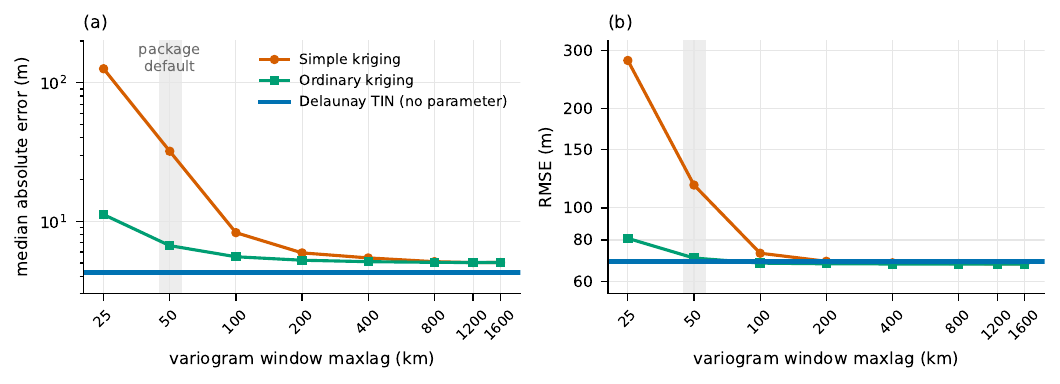}
\caption{
Sensitivity to variogram maxlag. The horizontal line is D-TIN, which has no variogram parameter. Kriging MedAE approaches the D-TIN reference as maxlag grows, while RMSE plateaus beyond 800~km.}
\label{fig:knob}
\end{figure*}

Fig.~\ref{fig:knob} shows that kriging is highly dependent on the variogram fitting window. At the 50~km setting, simple kriging has an RMSE of 117.3~m and a MedAE of 32~m (Table~\ref{tab:main}). Its RMSE varies by more than a factor of four across the range. Both simple and ordinary kriging approach a plateau beyond 800~km, so we use 800~km as a representative setting. D-TIN has no fitted scale parameter.

\subsection{Computational Cost}
\begin{figure}[t]
\centering
\includegraphics[width=0.92\columnwidth]{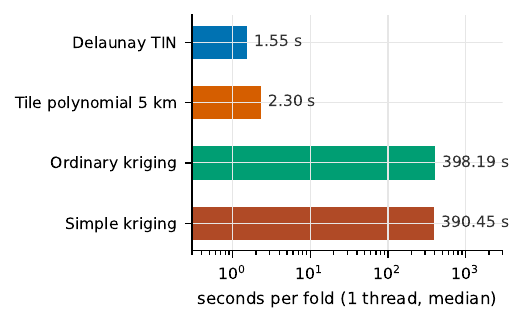}
\caption{Single-threaded runtime per fold. 
Values are the medians over 30 folds in 2005.}
\label{fig:time}
\end{figure}

The D-TIN workflows are substantially faster than the evaluated kriging workflows while achieving similar aggregate error at their best setting. D-TIN takes 1.5~s per fold and MS-D-TIN takes 1.1~s, compared with 390.4~s for SK800 and 398.2~s for OK800 (Table~\ref{tab:main} and Fig.~\ref{fig:time}). The median paired OK800-to-D-TIN ratio across the 30 folds is 252, though kriging also provides an uncertainty estimate based on fitted covariance.

\section{Ablation Study}
\label{sec:analysis}

\begin{table}[t]
\caption{Support and connectivity ablations. Near RMSE uses test observations within 0.5~km of training support. The means variants place cell averages at cell centers, and the nearest variants retain the real observation closest to each cell center. OK800 denotes ordinary kriging with an 800~km variogram window.  Points is the number of vertices supplied to the estimator. Points and runtime are fold medians.}

\label{tab:mechanism}
\centering
\scriptsize
\begin{tabular}{@{}lrrrr@{}}
\toprule
Representation & Points & RMSE/ m & Near RMSE / m & Time / s \\
\midrule
D-TIN & 176{,}953 & 68.8 & 6.1 & 1.5 \\
Per-tile D-TIN, 5 km & 176,953 & 89.2 & 7.8 & 1.7 \\
D-TIN, 1 km means & 46,866 & 68.7 & 11.0 & 0.6 \\
D-TIN, 1 km nearest & 46,866 & 69.0 & 7.9 & 0.6 \\
OK800, 1 km means & 46,866 & 67.7 & 8.3 & 398.2 \\
OK800, raw support & 176,953 & 67.9 & 5.5 & 676.0 \\
\bottomrule
\end{tabular}
\end{table}

\subsection{Cross-Boundary Connectivity}
The per-tile D-TIN restricts prediction to independent tiles while keeping every raw observation. It uses the same piecewise-linear interpolation inside each local convex hull. A query outside that hull uses the nearest observation in the tile, and an empty tile uses the globally nearest training observation. RMSE rises from 68.8 to 89.2~m (Table~\ref{tab:mechanism}).

The near-support RMSE increases by 1.7~m, while the aggregate RMSE increases by 20.4~m. Thus, the boundary restriction affects the aggregate error more strongly than the near-support error.

The tile polynomial also faces a rank problem because observations within one tile are often nearly collinear. The singular-value rule prevents unstable high-order fits, but it cannot recover support outside the partition. 
The similar RMSE of per-tile D-TIN (89.2 m) and the tile polynomial (89.5 m) indicates that restricting support at tile boundaries accounts for most of the difference between the unrestricted D-TIN and tile baseline (Tables~\ref{tab:mechanism} and \ref{tab:main}).

\subsection{Support Density and Placement} 

Table~\ref{tab:mechanism} tests whether D-TIN depends on retaining every raw observation. The two 1~km D-TIN variants have the same number of vertices. One places cell means at cell centers, while the other retains the real observation nearest each cell node. Their aggregate RMSE differs by 0.3~m, but near-support RMSE differs by 3.1~m. 
Because cell means change both location and elevation through averaging, these two effects remain coupled. Their joint effect is concentrated near observed support.

Comparison of D-TIN with the nearest-point variant shows the effect of thinning. Reducing the median vertex count from 176,953 to 46,866 changes aggregate RMSE from 68.8 to 69~m. Near RMSE increases from 6.1 to 7.9~m. So retaining every observation is not necessary for overall RMSE, but those removed vertices improve reconstruction close to observed support.

\subsection{Estimator and Representation}
Raw-support kriging provides the complementary control. Removing the grid lowers near-support RMSE from 8.3 to 5.5~m, while aggregate RMSE changes from 67.7 to 67.9~m. Runtime rises from 398.2 to 676~s (Table~\ref{tab:mechanism}). Retaining raw support improves local reconstruction for both estimators, while gridding reduces the computational cost for kriging.

Overall, the controls separate local support loss from the effect of restricted tile prediction. Thinning and grid aggregation mainly affect nearby queries. Tile boundaries also affect the predictions at larger support distances, which is shown in aggregate RMSE.

\section{Discussion}
\label{sec:discussion}
\subsection{Connectivity and Support Placement}

The ablations separate the effects of cross-boundary access from support aggregation. 
Fixed tiles remove cross-tile context, while gridding moves and averages the support. 
As shown in Section~\ref{sec:analysis}, restricting D-TIN to independent tiles nearly reproduces the tile-polynomial error, so the ranking cannot be explained by polynomial degree alone. 
By contrast, support aggregation has little effect on aggregate RMSE but changes accuracy near observations.

\subsection {Accuracy Depends on Support Distance}
Most held-out points are several kilometers from training support, where smoother prediction reduces some large residuals and improves RMSE. D-TIN has a lower MedAE and performs better near raw support, but a linear triangle cannot represent curvature within its interior. MS-D-TIN combines predictions from coarser supports and gives the best aggregate RMSE. Its slightly higher MedAE indicates that this gain does not improve every part of the residual distribution.

\subsection{Implications for Track-Sampled Reconstruction}
The evaluation protocol is part of the empirical claim. Continuous windows separate local along-track neighbors, and reporting error by distance shows what the aggregate hides. The same design may apply to other path-sampled measurements when window and buffer scales follow the sensor geometry.

For this task, D-TIN provides a direct reconstruction from irregular observations. It requires no grid resolution, covariance fitting window, or prediction neighborhood. It produces a continuous piecewise-linear surface inside the observed convex hull, and exposes a sparse adjacency graph for downstream analysis. MS-D-TIN reuses the same construction on coarsened supports. Gridded kriging remains appropriate when a common raster or covariance model is required. Fixed tiles remain useful for storage and scheduling, but prediction should share support across their boundaries.

\subsection{Computational Implications}
The two orders of magnitude runtime gap between OK800 and D-TIN (Table~\ref{tab:main}) changes which analyses are practical. Spatial cross-validation, sensitivity sweeps, and repeated campaign processing multiply the cost of each fit. D-TIN also avoids variogram fitting and repeated neighborhood solves. 
Its mesh can be reused when several query sets share the same training support.

\subsection{Limitations and Future work}
The proposed method and study address multiple outstanding challenges but have a few limitations we will address in future work.
Firstly, the proposed method is best used to estimate elevations inside the observed convex hull rather than transfer to an unseen region or mission. Extrapolation beyond the hull will need new strategies.

Secondly, the kriging results used here apply to the evaluated GStatSim workflow and do not cover nonstationary or anisotropic covariance models. A more complete comparison will need to add those. Finally, the scales and weights in the current multiscale variant are fixed, so the observed improvement does not establish an optimal combination of either or its mechanism. 

D-TIN also provides no native prediction uncertainty though any estimates on observation uncertainty carry over trivially.
The Delaunay mesh is the natural sparse spatial graph. The learned message-passing models could use this graph while adapting the fixed barycentric weights \cite{kipf2017,pfaff2021}. Future work can test learned and multiscale variants in denser missions such as ICESat-2.

\section{Conclusion}
Track-sampled altimetry combines dense along-track observations with sparse cross-track support. 
D-TIN applies the observed geometry directly to surface reconstruction. Across seven ICESat-1 campaigns, it consistently improves RMSE over the fixed 5~km tile polynomial, with the largest gains in steep terrain and across wide gaps. Restricting D-TIN to the same tile boundaries reproduces most of the tile boundary related error, which identifies the partitioning as a main factor in this source of error.
D-TIN and the evaluated OK800 workflow have similar aggregate error with different strengths. OK800 lowers RMSE by 1.1~m, while D-TIN lowers MedAE by 0.8~m. OK800 takes 252 times as long in the paired median runtime. MS-D-TIN lowers aggregate RMSE to 67.5~m while retaining low computational cost. 

The ablations show why representation matters. Thinning has little effect on aggregate RMSE, but local accuracy decreases when support is averaged, displaced, or restricted by fixed boundaries. For track-sampled reconstruction, support placement, connectivity, and measured computation should therefore be evaluated as model choices. The multiscale result also motivates extensions that preserve the D-TIN graph while combining information across support resolutions.

\section*{Acknowledgments}
This work was supported by the National Science Foundation under
grants 2050412 and 2004826, and by the Laidlaw Foundation. We acknowledge many fruitful discussions with Professor B. Csatho and her group at the University at Buffalo.

\section*{Data Availability}
ICESat-1 GLA12 Version~34 is publicly available from the National Snow and Ice Data Center \cite{zwally2014gla12}.

\bibliographystyle{IEEEtran}
\bibliography{references}

@article{bamber2018,
  author  = {Bamber, Jonathan L. and Westaway, Richard M. and Marzeion, Ben and Wouters, Bert},
  title   = {The land ice contribution to sea level during the satellite era},
  journal = {Environmental Research Letters},
  volume  = {13},
  pages   = {063008},
  year    = {2018},
  doi     = {10.1088/1748-9326/aac2f0}
}

@article{imbie2020,
  author  = {Shepherd, Andrew and Ivins, Erik and Rignot, Eric and Smith, Ben and van den Broeke, Michiel and others},
  title   = {Mass balance of the {Greenland} Ice Sheet from 1992 to 2018},
  journal = {Nature},
  volume  = {579},
  pages   = {233--239},
  year    = {2020},
  doi     = {10.1038/s41586-019-1855-2}
}

@article{zwally2002,
  author  = {Zwally, H. Jay and Schutz, Bob and Abdalati, Waleed and Abshire, James and Bentley, Charles and others},
  title   = {{ICESat}'s laser measurements of polar ice, atmosphere, ocean, and land},
  journal = {Journal of Geodynamics},
  volume  = {34},
  pages   = {405--445},
  year    = {2002},
  doi     = {10.1016/S0264-3707(02)00042-X}
}

@article{schutz2005,
  author  = {Schutz, B. E. and Zwally, H. J. and Shuman, C. A. and Hancock, D. and DiMarzio, J. P.},
  title   = {Overview of the {ICESat} mission},
  journal = {Geophysical Research Letters},
  volume  = {32},
  pages   = {L21S01},
  year    = {2005},
  doi     = {10.1029/2005GL024009}
}

@article{helm2014,
  author  = {Helm, V. and Humbert, A. and Miller, H.},
  title   = {Elevation and elevation change of {Greenland} and {Antarctica} derived from {CryoSat-2}},
  journal = {The Cryosphere},
  volume  = {8},
  pages   = {1539--1559},
  year    = {2014},
  doi     = {10.5194/tc-8-1539-2014}
}

@article{csatho2014,
  author  = {Csath{\'o}, B. M. and Schenk, A. F. and van der Veen, C. J. and Babonis, G. and Duncan, K. and others},
  title   = {Laser altimetry reveals complex pattern of {Greenland} Ice Sheet dynamics},
  journal = {Proceedings of the National Academy of Sciences},
  volume  = {111},
  pages   = {18478--18483},
  year    = {2014},
  doi     = {10.1073/pnas.1411680112}
}

@article{serac2012,
  author  = {Schenk, Tony and Csath{\'o}, Be{\'a}ta},
  title   = {A new methodology for detecting ice sheet surface elevation changes from laser altimetry data},
  journal = {IEEE Transactions on Geoscience and Remote Sensing},
  volume  = {50},
  number  = {9},
  pages   = {3302--3316},
  year    = {2012},
  doi     = {10.1109/TGRS.2011.2182357}
}

@article{alps2021,
  author  = {Shekhar, Prashant and Csath{\'o}, Be{\'a}ta and Schenk, Tony and Roberts, Carolyn and Patra, Abani K.},
  title   = {{ALPS}: A unified framework for modeling time series of land ice changes},
  journal = {IEEE Transactions on Geoscience and Remote Sensing},
  volume  = {59},
  number  = {8},
  pages   = {6466--6481},
  year    = {2021}
}

@article{mackie2023,
  author  = {MacKie, E. J. and Field, M. and Wang, L. and Yin, Z. and Schoedl, N. and Hibbs, M. and Zhang, A.},
  title   = {{GStatSim} v1.0: a {Python} package for geostatistical interpolation and conditional simulation},
  journal = {Geoscientific Model Development},
  volume  = {16},
  pages   = {3765--3783},
  year    = {2023},
  doi     = {10.5194/gmd-16-3765-2023}
}

@incollection{shewchuk1996,
  author    = {Shewchuk, Jonathan Richard},
  title     = {Triangle: Engineering a {2D} quality mesh generator and {Delaunay} triangulator},
  booktitle = {Applied Computational Geometry: Towards Geometric Engineering},
  series    = {Lecture Notes in Computer Science},
  volume    = {1148},
  publisher = {Springer},
  pages     = {203--222},
  year      = {1996},
  doi       = {10.1007/BFb0014497}
}

@article{rippa1990,
  author  = {Rippa, Shmuel},
  title   = {Minimal roughness property of the {Delaunay} triangulation},
  journal = {Computer Aided Geometric Design},
  volume  = {7},
  pages   = {489--497},
  year    = {1990}
}

@article{larour2012,
  author  = {Larour, E. and Seroussi, H. and Morlighem, M. and Rignot, E.},
  title   = {Continental scale, high order, high spatial resolution, ice sheet modeling using the {Ice Sheet System Model} ({ISSM})},
  journal = {Journal of Geophysical Research: Earth Surface},
  volume  = {117},
  pages   = {F01022},
  year    = {2012},
  doi     = {10.1029/2011JF002140}
}

@article{roberts2017,
  author  = {Roberts, David R. and Bahn, Volker and Ciuti, Simone and Boyce, Mark S. and Elith, Jane and others},
  title   = {Cross-validation strategies for data with temporal, spatial, hierarchical, or phylogenetic structure},
  journal = {Ecography},
  volume  = {40},
  pages   = {913--929},
  year    = {2017},
  doi     = {10.1111/ecog.02881}
}

@article{ploton2020,
  author  = {Ploton, Pierre and Mortier, Fr{\'e}d{\'e}ric and R{\'e}jou-M{\'e}chain, Maxime and Barbier, Nicolas and Picard, Nicolas and others},
  title   = {Spatial validation reveals poor predictive performance of large-scale ecological mapping models},
  journal = {Nature Communications},
  volume  = {11},
  pages   = {4540},
  year    = {2020},
  doi     = {10.1038/s41467-020-18321-y}
}

@inproceedings{kipf2017,
  author    = {Kipf, Thomas N. and Welling, Max},
  title     = {Semi-supervised classification with graph convolutional networks},
  booktitle = {International Conference on Learning Representations},
  year      = {2017}
}

@inproceedings{pfaff2021,
  author    = {Pfaff, Tobias and Fortunato, Meire and Sanchez-Gonzalez, Alvaro and Battaglia, Peter W.},
  title     = {Learning mesh-based simulation with graph networks},
  booktitle = {International Conference on Learning Representations},
  year      = {2021}
}

@article{shekharpatra2020,
  author  = {Shekhar, Prashant and Patra, Abani},
  title   = {Hierarchical approximations for data reduction and learning at multiple scales},
  journal = {Foundations of Data Science},
  volume  = {2},
  pages   = {123--154},
  year    = {2020},
  doi     = {10.3934/fods.2020008}
}

@article{virtanen2020,
  author  = {Virtanen, Pauli and Gommers, Ralf and Oliphant, Travis E. and Haberland, Matt and Reddy, Tyler and others},
  title   = {{SciPy} 1.0: fundamental algorithms for scientific computing in {Python}},
  journal = {Nature Methods},
  volume  = {17},
  pages   = {261--272},
  year    = {2020},
  doi     = {10.1038/s41592-019-0686-2}
}

@article{chaplot2006,
  author  = {Chaplot, Vincent and Darboux, Fr{\'e}d{\'e}ric and Bourennane, Hocine and Legu{\'e}dois, Sophie and Silvera, Norbert and Phachomphon, Konngkeo},
  title   = {Accuracy of interpolation techniques for the derivation of digital elevation models in relation to landform types and data density},
  journal = {Geomorphology},
  volume  = {77},
  number  = {1--2},
  pages   = {126--141},
  year    = {2006},
  doi     = {10.1016/j.geomorph.2005.12.010}
}

@article{felikson2017,
  author  = {Felikson, Denis and Urban, Timothy J. and Gunter, Brian C. and others},
  title   = {Comparison of elevation change detection methods from {ICESat} altimetry over the {Greenland} Ice Sheet},
  journal = {IEEE Transactions on Geoscience and Remote Sensing},
  volume  = {55},
  year    = {2017},
  doi     = {10.1109/TGRS.2017.2709303}
}

@article{hohle2009,
  author  = {H{\"o}hle, Joachim and H{\"o}hle, Michael},
  title   = {Accuracy assessment of digital elevation models by means of robust statistical methods},
  journal = {ISPRS Journal of Photogrammetry and Remote Sensing},
  volume  = {64},
  number  = {4},
  pages   = {398--406},
  year    = {2009},
  doi     = {10.1016/j.isprsjprs.2009.02.003}
}

@misc{zwally2014gla12,
  author       = {Zwally, H. Jay and Schutz, Robert and Bentley, Charles and Bufton, John and Herring, Thomas and Minster, Jean-Bernard and Spinhirne, James and Thomas, Robert},
  title        = {{GLAS/ICESat L2 Antarctic and Greenland Ice Sheet Altimetry Data (GLA12), Version 34}},
  year         = {2014},
  howpublished = {[Data Set]. Boulder, Colorado {USA}: {NASA} National Snow and
                  Ice Data Center Distributed Active Archive Center},
  doi          = {10.5067/ICESAT/GLAS/DATA225}
}

@article{morlighem2017,
  author  = {Morlighem, M. and Williams, C. N. and Rignot, E. and An, L. and Arndt, J. E. and Bamber, J. L. and others},
  title   = {{BedMachine} v3: Complete bed topography and ocean bathymetry mapping of {Greenland} from multibeam echo sounding combined with mass conservation},
  journal = {Geophysical Research Letters},
  volume  = {44},
  number  = {21},
  pages   = {11051--11061},
  year    = {2017},
  doi     = {10.1002/2017GL074954}
}

@incollection{lawson1977,
  author    = {Lawson, Charles L.},
  title     = {Software for {$C^1$} Surface Interpolation},
  booktitle = {Mathematical Software {III}},
  editor    = {Rice, John R.},
  publisher = {Academic Press},
  address   = {New York, NY},
  pages     = {161--194},
  year      = {1977}
}

@article{slater2018,
  author  = {Slater, Thomas and Shepherd, Andrew and McMillan, Malcolm and Muir, Alan and Gilbert, Lin and Hogg, Anna E. and Konrad, Hannes and Parrinello, Tommaso},
  title   = {A New Digital Elevation Model of {Antarctica} Derived from {CryoSat-2} Altimetry},
  journal = {The Cryosphere},
  volume  = {12},
  number  = {4},
  pages   = {1551--1562},
  year    = {2018}
}

@book{cressie1993,
  author    = {Cressie, Noel A. C.},
  title     = {Statistics for Spatial Data},
  edition   = {Revised},
  publisher = {John Wiley \& Sons},
  address   = {New York, NY},
  year      = {1993}
}

\end{document}